\documentclass[aps,twocolumn,showpacs,groupedaddress]{revtex4}  
\usepackage{amsmath}
\usepackage{amssymb}
\usepackage{graphicx}
\usepackage{tikz}
\usepackage{pgfplots}

\begin{document}
\title{Quantum control of $^{88}$Sr$^+$ in a miniature linear Paul trap}
\author{N. Akerman, S. Kotler, Y. Glickman, A. Keselman and R.
Ozeri\\
\small \emph{Physics of Complex Systems, Weizmann Institute of
Science, Rehovot 76100, Israel} }
\date{\today}

\begin{abstract}
We report on the construction and characterization of an apparatus for quantum information experiments using $^{88}$Sr$^+$ ions. A miniature linear radio-frequency (rf) Paul trap was designed and built. Trap frequencies above $1$ MHz in all directions are obtained with $50$ V on the trap end-caps and less than $1$ W of rf power. We encode a quantum bit (qubit) in the two spin states of the $S_{1/2}$ electronic ground-state of the ion. We constructed all the necessary laser sources for laser cooling and full coherent manipulation of the ions' external and internal states. Oscillating magnetic fields are used for coherent spin rotations. High-fidelity readout as well as a coherence time of $2.5$ ms are demonstrated. Following resolved sideband cooling the average axial vibrational quanta of a single trapped ion is $\bar n=0.05$ and a heating rate of $\dot{\bar n}=0.016$ ms$^{-1}$ is measured.
\end{abstract}
\maketitle

\newpage

\section{Introduction}
\label{intro}

Trapped ions have proven to be a valuable resource in the field of experimental quantum information and precision measurements. Their tight spatial localization, isolation from
external perturbations and the ability of experimenters to coherently manipulate their internal and external degrees of freedom with high fidelity, place trapped ions among the leading technologies of quantum information research \cite{Review, Brief_Review}.

The realization of scalable quantum computation and simulation requires the design and fabrication of small trap arrays. The smallest traps demonstrated to date were built using modern micro-fabrication techniques \cite{micro_fabrication_trap,Signe_flat_trap,Lucent_trap,Joe_Si_trap,Sandia_trap,Superconducting_trap,heating_rate_haffner,Winnie_review}. Due to the technical complexity of micro-fabrication, most of the current research in this field is still conducted using manually assembled, mm-scale, traps. Here we report on the manual assembly of a miniature, sub mm-scale, trap. Our trap has dimensions that approach those of micro-fabricated traps but its construction required only commercial parts, conventional workshop capabilities and laser machining which is also commercially available.

Various internal states can be used to encode a qubit, examples include optical, hyperfine and Zeeman splitted levels \cite{hyperfineQubit1,hyperfineQubit2,ZeemanCa40,opticalQubit}. Here we focus on a Zeeman qubit in a $^{88}$Sr$^+$ ion \cite{Sinclair_Srqubit, Chuang_Srqubit}. We have constructed all the frequency stabilized lasers, that drive the various transitions required for laser-cooling, state-selective fluorescence detection, and coherent control. One convenient property of the Strontium ion is that all its relevant optical transition wavelengths are available using diode lasers.

Using a narrow line-width laser, we perform electron shelving on a narrow quadrupole transition, and are thus able to demonstrate high fidelity qubit readout. With the same laser we performed resolved sideband cooling and demonstrated $95\%$ ground state occupation of the axial vibrational mode. A heating rate of $\dot{\bar n}=0.016$ ms$^{-1}$ for this mode is measured.

Although a ground-state Zeeman qubit has practically an infinite life time, a coherent superposition is hard to maintain in the presence of ambient magnetic field noise. We constructed a servo system that reduces magnetic field noise and which leads to the extension of the qubit coherence time by an order of magnitude to $2.5$ ms. In this paper we give an overview of the apparatus we built and discuss various aspects of its performance.

\section{Experimental Apparatus}
\label{Apparatus}

\subsection{Trap and vacuum system}

\begin{figure}[h]
\center
\begin{tikzpicture}
   \begin{axis}[scale only axis,name=a,width=7cm,height=3.5cm,hide x axis,hide y axis]
   \addplot graphics[xmin=0,xmax=1,ymin=0,ymax=1] {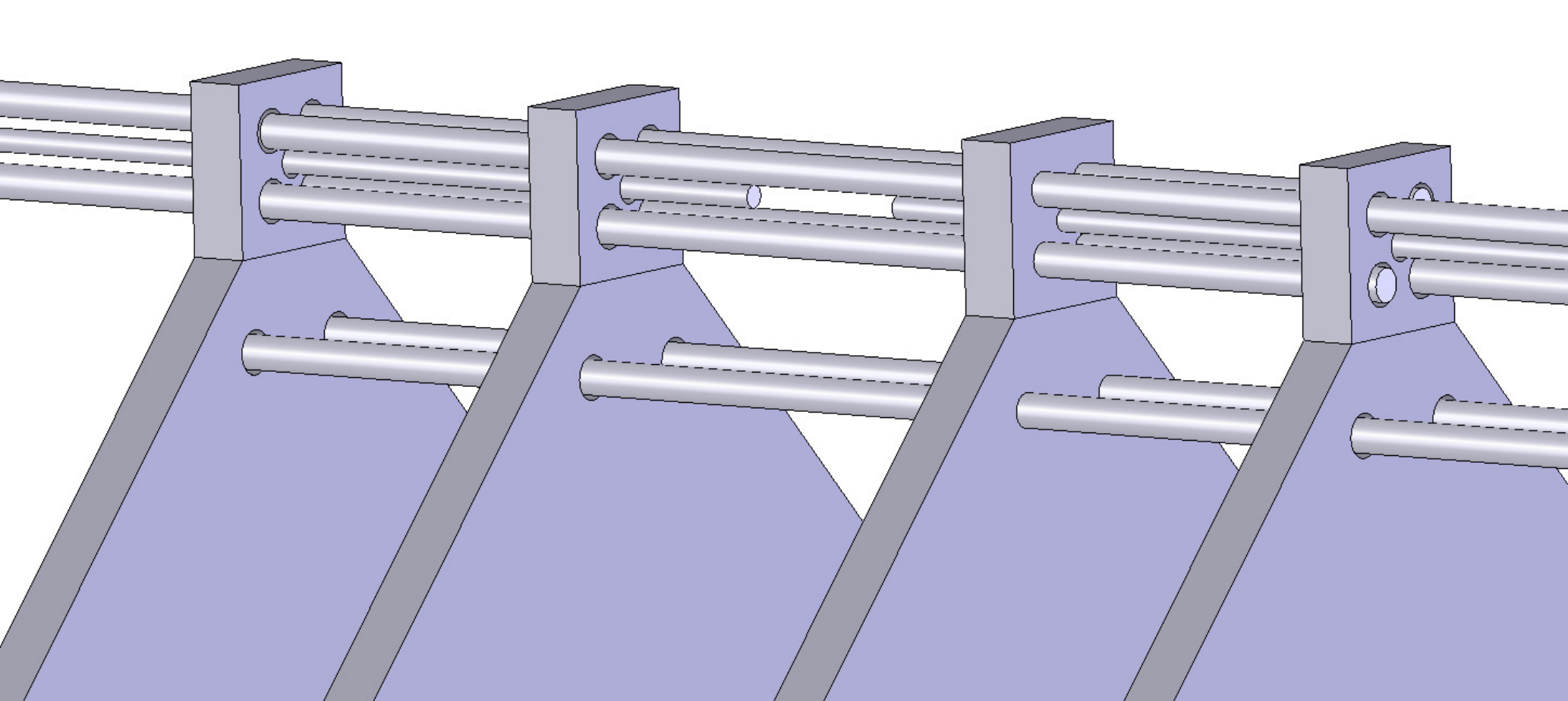};
   \end{axis}
   \draw (current axis.outer north west) node[anchor=north west] {(a)};
   \begin{axis}[scale only axis,at=(a.south west),width=7cm,height=4.55cm,hide x axis,hide y axis,anchor=north west]
   \addplot graphics[xmin=0,xmax=1,ymin=0,ymax=1] {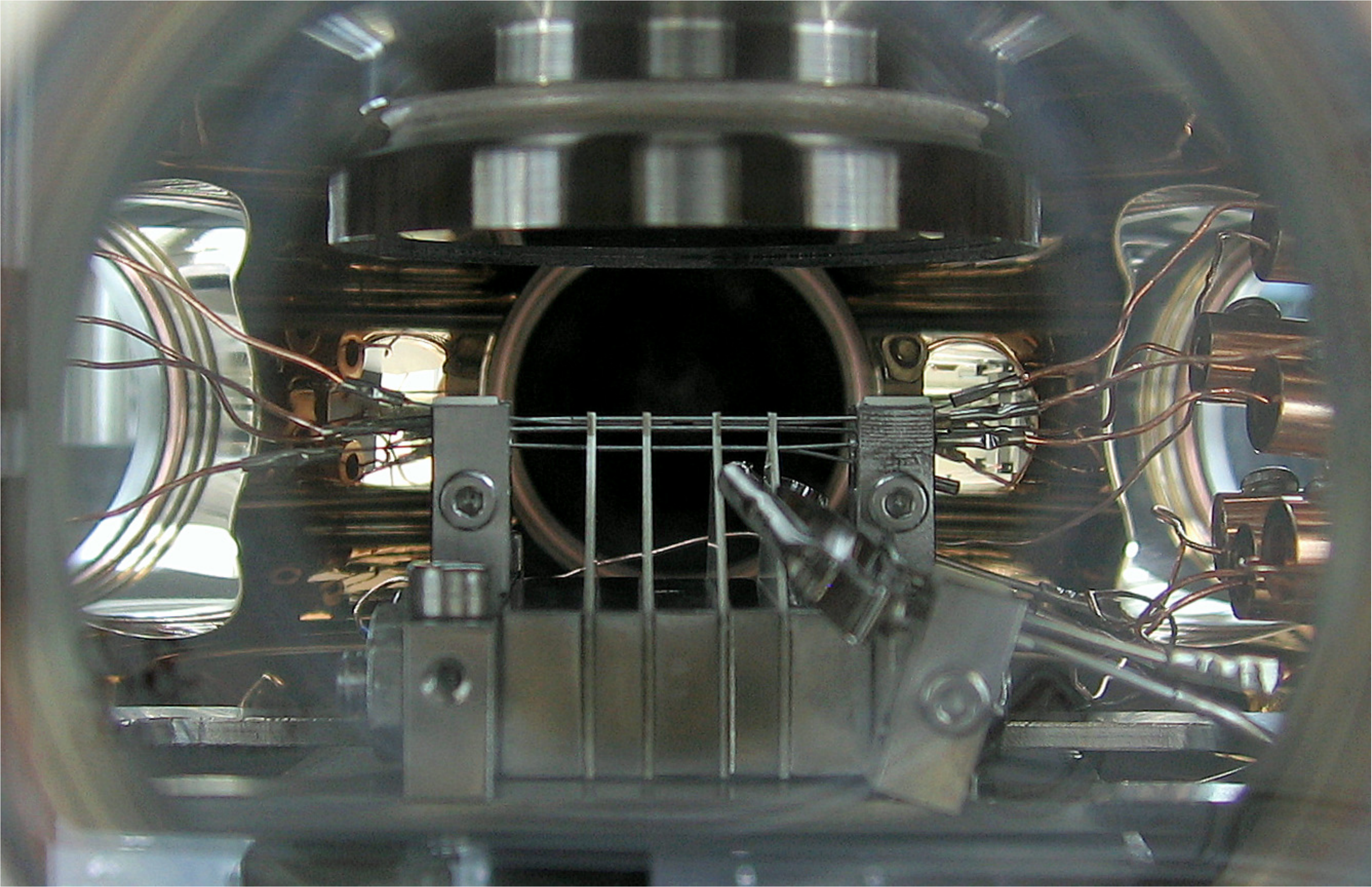};
   \end{axis}
   \draw (current axis.outer north west) node[anchor=north west] {(b)};
\end{tikzpicture}
\caption{(a) A close-up drawing of our linear Paul trap. The trap is constructed of six tungsten rods, held in place by four laser-machined alumina wafers. Two additional rod-electrodes are placed below the trap. One electrode is used for stray field compensation while the other is used for driving an oscillating current to induce magnetic-dipole transitions. (b) An image of the trap mounted inside the vacuum chamber. The trap is held in place by a titanium construction. Two Sr ovens, one from each side of the trap are anchored to the titanium construction. One of these, pointing to the trap center, is seen on the right side of (b)}\label{trap zoom}
\end{figure}

Our trap has the generic linear Paul trap configuration of four parallel rods positioned at the corners of a square, and two additional ``end-cap'' rods that are aligned along the square center, symmetrically from each side of the trap; see Fig.\ref{trap zoom}(a). The four parallel rods carry the rf and dc voltages necessary for radial confinement, whereas the dc voltages on the end-caps provide confinement in the axial direction. The square sides length is $0.6$ mm and the distance between the two end-caps is $1.3$ mm. The four rods and two end-caps are made of tungsten with diameters of $0.3$ mm and $0.2$ mm respectively. Tungsten has several advantages over stainless steel. Most importantly, it is non-magnetic even after machining. Tungsten is also more rigid; an important property when working with small diameter rods. Moreover, Tungsten is superior in its electrical and thermal conductivity. The distances of the ion from the rf/dc electrodes and end-caps is $0.27$ mm and $0.65$ mm respectively.

Structural integrity and electric insulation are provided by four laser machined alumina wafers with adequate holes for holding the electrodes. These alumina wafers are aligned and held together by two, $5$ mm diameter, stainless steel rods that provide a skeleton for the trap structure. The wafers are separated by three titanium spacers of $3,\ 4,\ 3$ mm thickness. Here, Titanium is used because it is non-magnetic. The wafers thus define four straight support points for each electrode. Small naturally occurring misalignment causes mechanical friction that keeps the electrodes in place once they have been threaded through the four alumina wafers.

Two additional electrodes are placed below the trap, at a distance of $1.66$ mm from the ion. One electrode is used to drive current that produces the oscillating magnetic field needed for magnetic-dipole coupling between the qubit levels. The second electrode is used for compensation of stray electric fields in the direction of the rf electrodes.

The trap is driven at $\Omega_{rf}/2\pi=21$ MHz with $0.5$ W of power. A helical resonator with a Q-factor of $70$ produces a voltage gain to $200$ V on the rf electrodes.
The resulting radial trap frequencies are $\omega_{rad}^{1,2}/2\pi=\{2.5,2.35\}$ MHz. The degeneracy between the two radial directions is lifted by a $0.5$ V bias on the dc electrodes. The trap axial frequency is $\omega_{ax}/2\pi=1$ MHz with a voltage of $50$ V applied to the end-cap electrodes. The deviation of the trapping potential from a pure harmonic potential has been characterized as well. The next term beyond linear in the force along the axial direction is cubic, $F_{nl}=\alpha x^3$. This term is measured by the study of the ion non-linear response to a strong, near resonance, drive. The cubic force term coefficient is found to be $\alpha/(2\pi)^2=1.74\pm0.03\cdot10^{-7}$ KgHz$^2$/m$^2$ ~\cite{duffing}

The trap is mounted inside an octagon-shaped vacuum chamber (Kimball physics MCF600-SphOct-F2C8) with eight $2.75$ in. view-ports on the octagon faces and two $6$ in. ports at the top and the bottom of the chamber. The side ports are used for laser access,  electrical feedthrough and connection to the vacuum pumps whereas the top port is designated for imaging. Three types of vacuum pumps are combined to achieve ultra-high vacuum. A $20$ l/s  ion getter pump and a Titanium sublimation pump are connected to the main chamber through a $2.75$ in. port. Inside the main chamber we placed two non-evaporable getter strips (SAES St707) which are thermally ($300~^{\circ}$C) activated during bake-out. After bake-out, the vacuum ion-gauge was indicating a pressure below its baseline reading of $4\cdot10^{-12}$ mbar. During the last two years ions were regularly trapped for periods of up to a week. We never observed chemistry of trapped-ions with residual background molecules or loss of ions from the trap, as long as they were laser-cooled.

A thermal jet of neutral Strontium atoms is directed into the trap center from either one of two Sr ovens that are positioned approximately $2$ cm from the trap center. The ovens are made of $1$ mm diameter stainless-steel tubes with one end crimped. Small grains of Strontium metal are placed inside the tubes in a dry nitrogen atmosphere. We seal the ovens with a thin layer of Indium before we expose them to room atmosphere. Resistive heating the oven with a current of $2$ A, results in the jet of neutral Strontium atoms necessary for trap loading.

\subsection{The $^{88}$Sr$^+$ energy levels and laser sources}
A single or few ions are loaded into the trap by photo-ionization. Photo-ionization is performed using a resonant two-photon process \cite{Berkeland_photoionization, Sinclair_photoionization}. A Sr atom is initially excited to the $5s5p^1P_1$ level and then, by a second photon, to the auto-ionizing $5p^{2 1}D_2$ level, at transition wavelengths of $461$ nm and $405$ nm respectively. The first is generated by doubling a $922$ nm external-cavity diode laser (ECDL) in a single pass through a PPKTP crystal (Raicol Crystals Ltd.). The crystal is mounted in a thermally stabilized housing. A $100$ mW of $922$ nm light generates an output power of $100 \mu$W at $461$ nm. The second beam of $20$ mW at $405$ nm, is produced using a free running laser diode (InGaN). The $461$ nm ($405$ nm) beam is focused, at the trap center, to a waist of $50 \mu$m with a typical power of $15 \mu$W ($3$ mW).

\begin{figure}[h]
\center
\begin{tikzpicture}
   \begin{axis}[scale only axis,name=a,width=8cm,height=6cm,hide x axis,hide y axis]
   \addplot graphics[xmin=0,xmax=1,ymin=0,ymax=1] {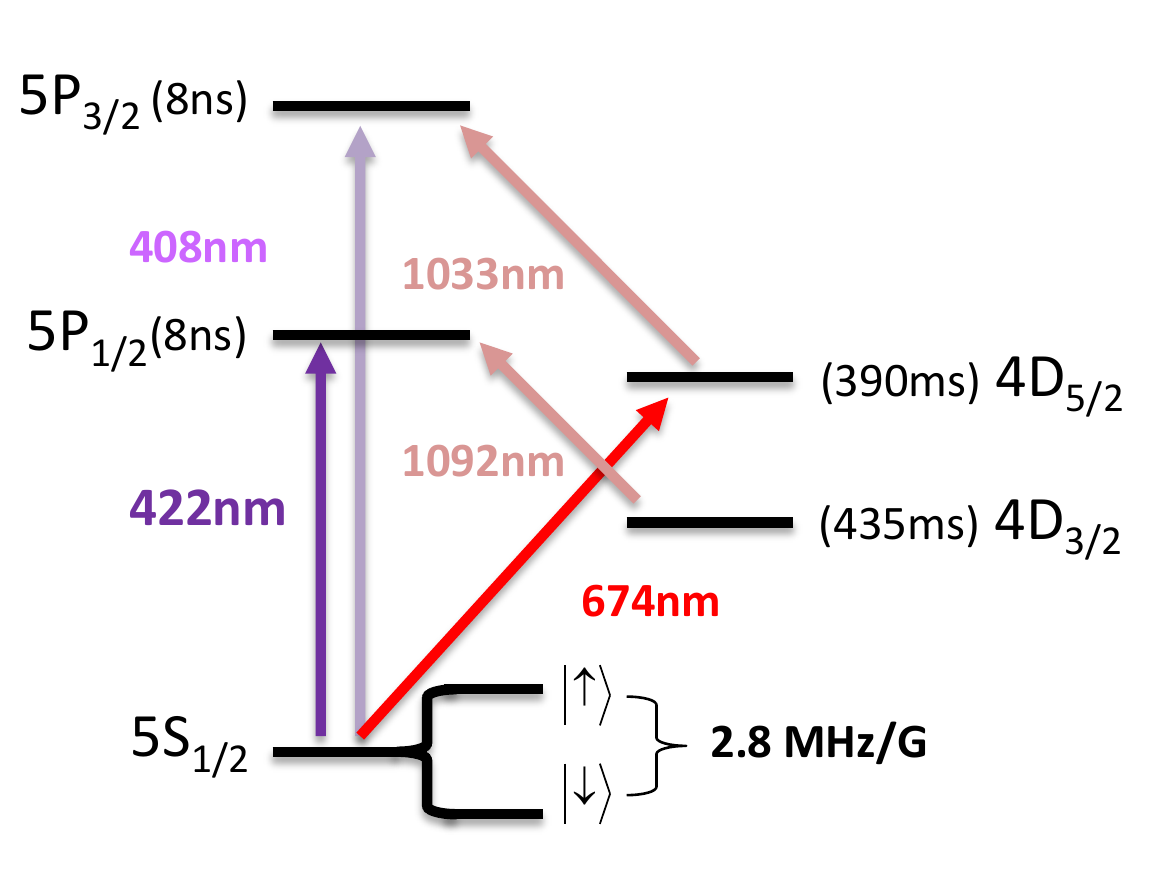};
   \end{axis}
   \draw (current axis.outer north west) node[anchor=north west] {(a)};
   \begin{axis}[scale only axis,at=(a.south west),width=8cm,height=6cm,hide x axis,hide y axis,anchor=north west]
   \addplot graphics[xmin=0,xmax=1,ymin=0,ymax=1] {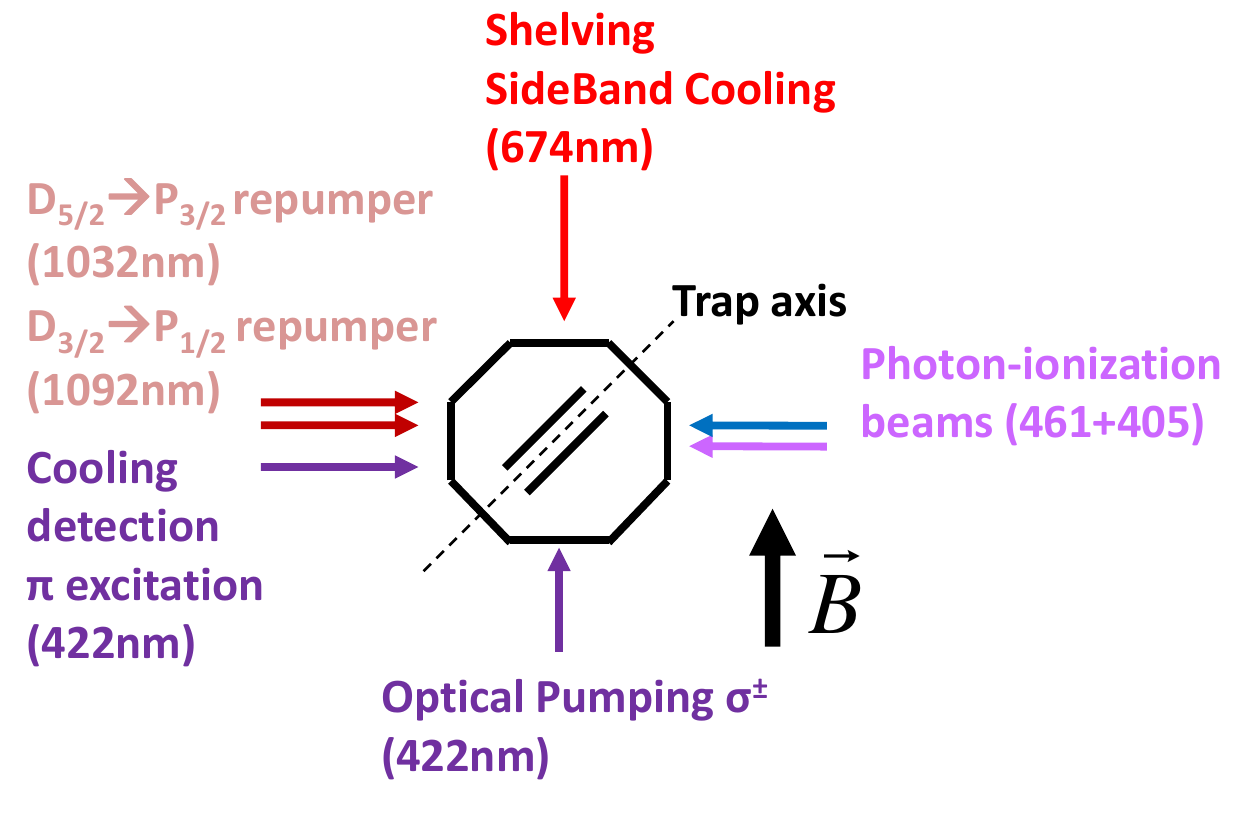};
   \end{axis}
   \draw (current axis.outer north west) node[anchor=north west] {(b)};
\end{tikzpicture}
\caption{(a) The $^{88}$Sr$^+$ ion energy levels diagram and relevant transition (in parenthesis are the levels life time). (b) Geometric arrangement of the various laser beams with respect to the trap long axis and the quantization axis(=magnetic field direction).}
\label{SrEnergyLevels}
\end{figure}

Doppler-cooling and florescence detection are preformed on the $S_{1/2} \rightarrow P_{1/2}$ electric-dipole transition at $422$ nm. See Fig.\ref{SrEnergyLevels}(a) for an energy-level diagram. Since Sr$^{+}$ has low laying $4D$ levels, the cooling cycle has to be closed with additional light at $1092$ nm which repumps population from the long-lived ($0.37$ s) $D_{3/2}$ level. The cooling light is generated by a $844$ nm ECDL which is frequency doubled, producing $15$ mW of light at $422$ nm (Toptica SHG110). The fundamental diode laser at $844$ nm is locked, using a Pound-Drever-Hall (PDH) error signal, to a hemispherical Fabry-Perot cavity . The cavity is built using an Zerodur spacer and has one of its mirrors mounted on a piezo-electric stack. The cavity maintains long-term stability by locking it to the $5s^2S_{1/2}\rightarrow 6p^2P_{1/2}$ transition in $^{85}$Rb which is $400$ MHz red detuned from the cooling transition \cite{Rb_reference}. The error signal is generated by passing $422$ nm light through a heated ($120$ C) Rb vapor cell and recording the derivative of a saturation absorption signal.  The repump light at $1092$ nm is generated by a distributed feedback (DFB) diode laser (Toptica DL100 DFB) producing $15$ mW. Another ECDL (Toptica DL100) at $1033$ nm with $30$ mW is needed to repump the $D_{5/2}$ level. Both the $1092$ nm and the $1033$ nm lasers are simultaneously locked to an in-vacuum Zerodur cavity with a finesse of $100$ using PDH and lock-in techniques respectively. Typical frequency drifts of the lasers due to thermal fluctuations of the cavity are of the order of $1$ MHz in few hours.

The narrow $S_{1/2}\rightarrow D_{5/2}$ quadrupole transition is driven by another ECDL (Toptica DL100) emitting $10$ mW at $674$ nm. The cavity of this laser is $10$ cm long leading to a fast line-width of approximately $100$ kHz in $10 \mu$s, significantly above the expected Schawlow-Townes limit. The ECDL is then locked to in-vacuum ($10^{-8}$ mbar) Ultra Low Expansion glass (ULE) cavity with a finesse of $100,000$ (ATFilms, \cite{JunYe_cavity}) using a PDH technique. A laser line-width of less than $80$ Hz is measured by preforming Ramsey experiment on the ion. Thermal fluctuations cause the cavity resonance to drift on a time scale of few hours with peak to peak amplitude of $1$ MHz (The lab temperature is stabilized to within $0.3~^{\circ}$C), with a typical slope of $10$ Hz/s. These slow drifts are automatically compensated for by scanning the atomic transition every few minutes and continuously updating the laser frequency by linear extrapolation. Using this technique, the laser detuning from any desired transition, can be smaller than $300$ Hz at all times.

All light sources pass through acousto-optical modulators (AOM's), in a double-pass configuration, before they are brought to the trap by polarization maintaining single mode fibers. The AOMs have two roles. First, they bridge the frequency difference between the frequency references and the atomic transitions. Second, they allow for switching and control of the light intensity by controlling their driving rf power. The frequency sources for the AOM's are based on either voltage controlled oscillators (VCO's) or direct digital synthesizers (DDS's) which are controlled by a field programable gate array (FPGA). The use of DDS's allows for phase and frequency control during a single experimental sequence.

The beams are focused at the trap center to a waist of $50 \mu$m. The schematic drawing of the laser beams configuration is presented in Fig.\ref{SrEnergyLevels}b. A magnetic field of $1-20$ G defines a quantization axis. The cooling and detection beams at $422$ nm are combined with the two repumping beams at $1092$ nm and $1033$ nm on a dichroic beam combiner before being focused on the ion. The beams are oriented at $45^{\circ}$ to the trap axis and perpendicular to the quantization axis while having a projection on both of the trap radial principle directions as well. A circularly polarized beam at $422$ nm, aligned parallel to the magnetic field, is used for optical pumping. The $674$ nm beam, that drives the $S_{1/2}\rightarrow D_{5/2}$ quadrupole transition, is also parallel the quantization axis. For this configuration, quadrupole transitions with $\Delta m=\pm1$ are allowed \cite{quandrupole_transition} with a Lamb-Dicke parameter of $\eta_{674}=0.05$.

\subsection{Imaging system and photons counting}
Ion fluorescence at $422$ nm is collected with an $N.A.\ 0.31$  objective lens (LENS-Optics). This lens was custom designed, based on \cite{alt_objective}, to correct for the view-port spherical aberrations. A flipping mirror switches between forming an image of the trapped-ions onto a electron multiplying CCD (EMCCD) camera(Andor Luca), with magnification of $42$ and a diffraction limited resolution ($0.8 \mu$m), or single photon counting by two photomultiplier tubes (PMT) each on a different side of a polarizing beam splitter (PBS) cube. Computer-controlled quarter and half wavelength retardation plates are positioned in front of the PBS to allow for full polarization analysis of fluorescence photons.

We measured the total photon detection efficiency by counting detection events in the process of single photon scattering. Single photon scattering is achieved by first applying a cooling pulse on the $S_{1/2}\rightarrow P_{1/2}$ transition. After an average of $14$ scattering events the electron decays to the meta-stable $D_{3/2}$ level. Then, a short pulse of $1092$ nm light pumps the electron back to the $P_{1/2}$ state, from which it decays back to the $S_{1/2}$ state via single photon emission. The total photon detection efficiency was thus found to be $2.5\cdot10^{-3}$. When both the $422$ nm and $1092$ nm beams are applied to the ion, the maximal photon detection rate is $70$ kHz. In the absence of an ion we measure a $1$ kHz photon detection rate due to scattering of laser light from trap surfaces.

\begin{figure}[h]
\center
\begin{tikzpicture}
\center
\begin{axis}[scale only axis,name=a,width=8cm,height=2.8cm,hide x axis,hide y axis,xmin=0,xmax=1,ymin=0,ymax=1]
\addplot graphics[xmin=0.15,xmax=0.9,ymin=0,ymax=1]{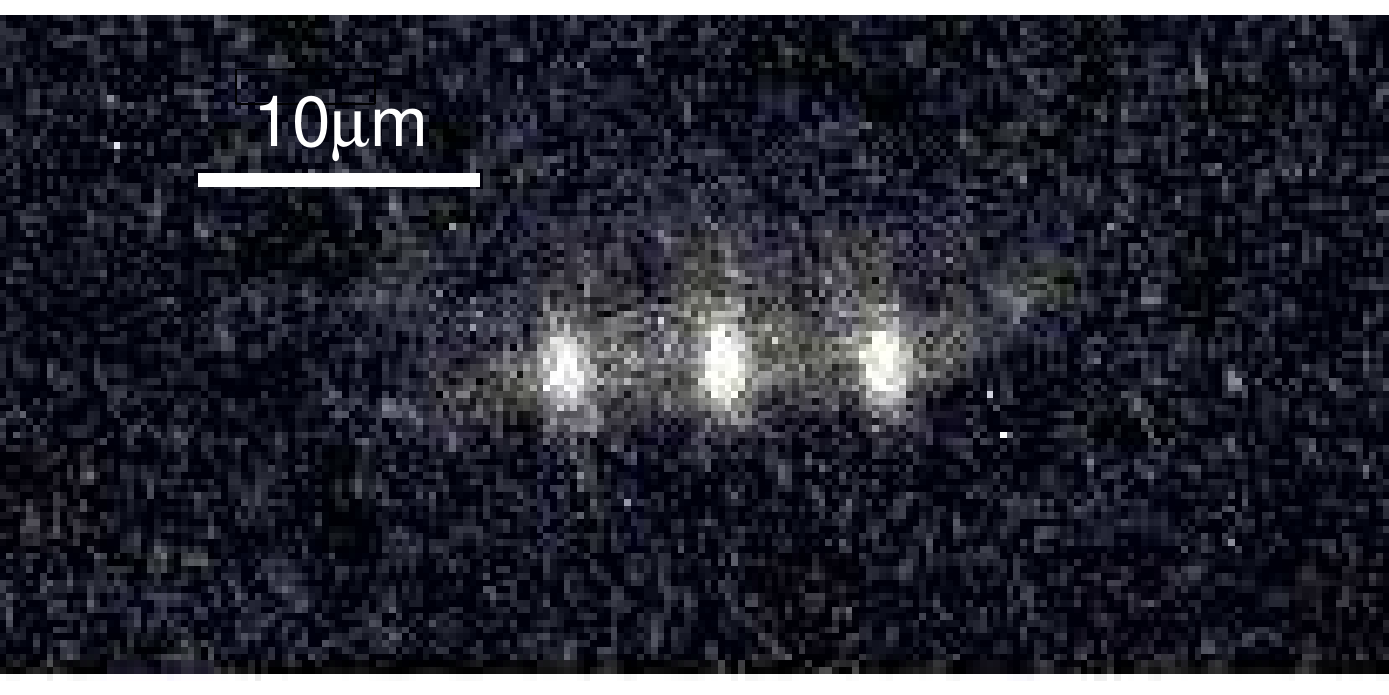};
\end{axis}
\draw (current axis.outer north west) node[anchor=north west] {(a)};
\begin{axis}[scale only axis,at=(a.south west),width=8cm,height=6cm,hide x axis,hide y axis,anchor=north west]
\addplot graphics[xmin=0,xmax=1,ymin=0,ymax=1,includegraphics={trim=12 9 12 8,clip}] {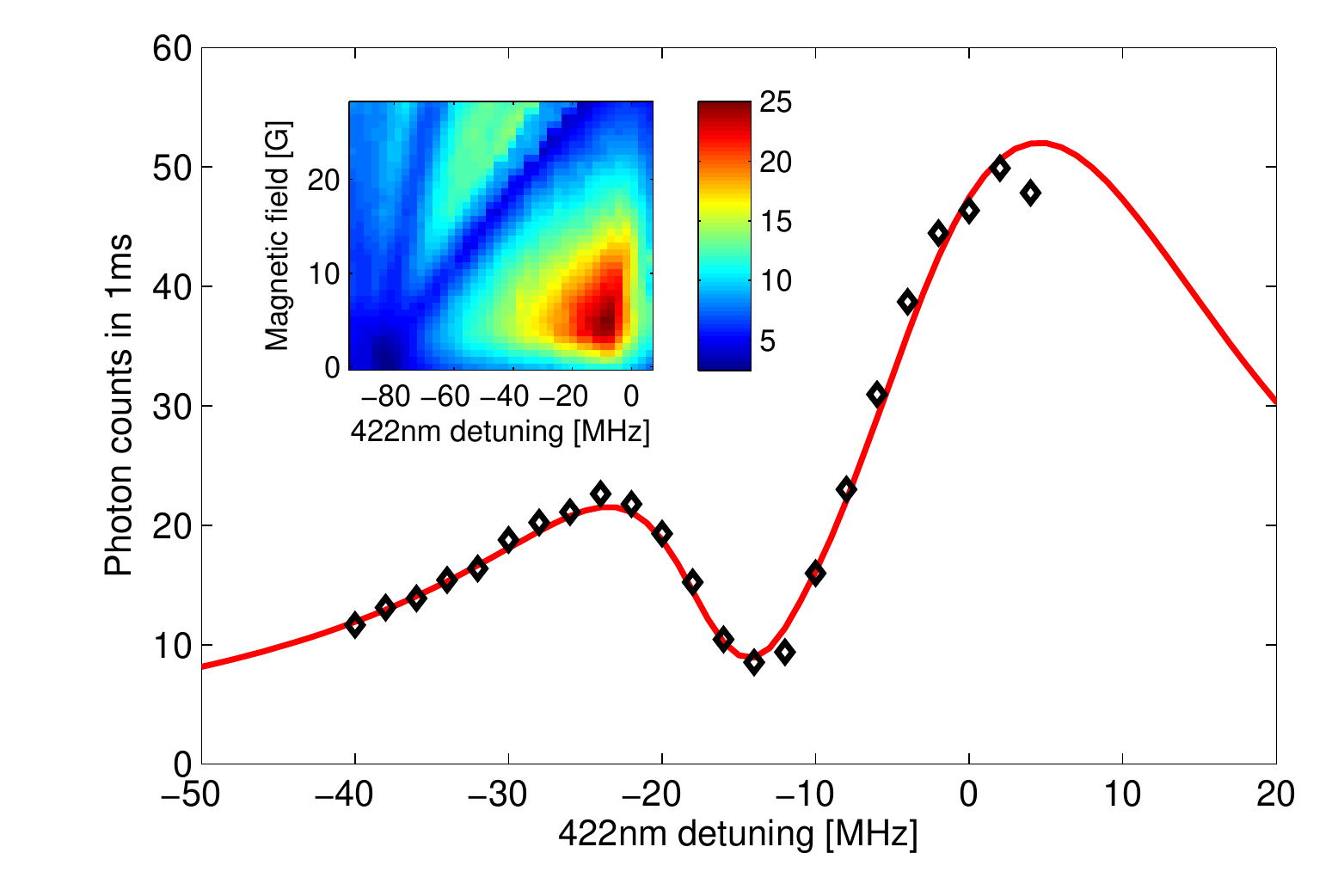};
\end{axis}
\draw (current axis.outer north west) node[anchor=north west] {(b)};
\end{tikzpicture}
\caption{(a) An EMCCD image of a three ion crystal. (b) Spectroscopic scan of the $S_{1/2}-P_{1/2}$ transition. Single ion fluorescence counts in 1 ms as function of the $422$ nm laser detuning $\Delta_{422}$ in the present of the $1092$ nm repump. The red line is a fit based on the solution of an eight-level coupled optical Bloch equations\cite{darkResonance}. The fit parameters are the repump laser detuning $\Delta_{1092}=-14$ MHz, the laser intensities $I/I_{sat\ 1092}=7$, $I/I_{sat\ 422}=0.6$, the magnetic field on the ion, $B=1.18$ G, and the laser linewidths, $\gamma_L=0.5$ MHz. The $422$ nm  and $1092$ nm beams polarizations are parallel and perpendicular to the magnetic field respectively. The inset shows a 2D scan of fluorescence as function of the $422$ nm laser detuning, $\Delta_{422}$ and the magnetic field. Here the $1092$ nm repump is detuned by $\Delta_{1092}=-80$ MHz from resonance. As the magnetic field is increased the dark resonances due to different Zeeman sub-levels separate and can be resolved in the fluorescence spectrum.}
\label{IonPic_SPline}
\end{figure}

Figure \ref{IonPic_SPline}(a) shows an EMCCD image of a three ion crystal with inter-ion separation of $5.5 \mu$m. Fig.\ref{IonPic_SPline}(b) shows fluorescence spectroscopy of the $S_{1/2}\rightarrow P_{1/2}$ transition based on PMT photons counting. The dip on the left side of the spectrum is a dark resonance resulting from transition amplitude interference in the presence of both the $422$ nm and the $1092$ nm light. The dependence of the spectrum on the magnetic field is shown in the inset. As the magnetic field is increased, the dark resonances due to different Zeeman sub-levels separate and can be resolved.

\section{Micro-motion compensation}
\label{secMicromotion}
\begin{figure}[h]
\center
\includegraphics[width=8cm]{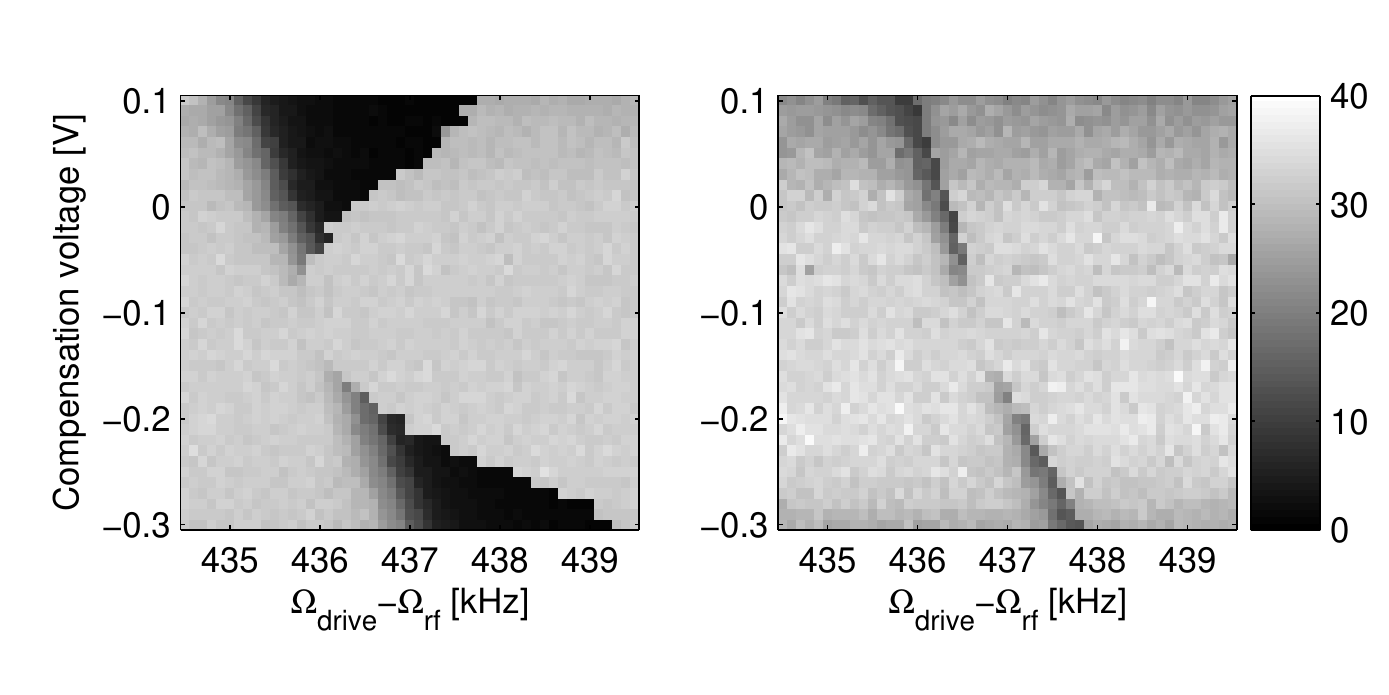}
\caption{Micro-motion compensation scan. Fluorescence counts as function of compensation voltage on the DC electrode pair and drive frequency for both increasing (left) and decreasing (right) frequency sweeps. The fluorescence drops (shows as dark region) when the ion experience resonance drive. The ion is brought to the rf minimum at -0.1V for which no fluorescence drop is seen. The asymmetry and hysteretic behavior are a result of the nonlinearity of the trapping potential.}
\label{figmicromotion}
\end{figure}

Displacement of the ion from the rf potential minimum due to stray electric fields or structural deformations causes excess micro-motion, i.e. oscillation of the ion at the rf trap frequency, which deteriorates cooling and detection efficiency. Once the excess micro-motion is probed it can be eliminated by proper voltage biasing of different trap electrodes in order to bring the ion to the rf minimum. There are various methods for probing micro-motion\cite{micromotion_methods}. We implement a method that was recently demonstrated by the NIST ion storage group \cite{Private_communication}. We inject a small additional rf voltage at a frequency $\Omega_{drive}=\Omega_{rf}+\omega_{ax}$ to the trap rf resonator. Due to potential nonlinearities, a drive at the trap frequency, $\omega_{ax}$, is produced with an amplitude that nulls at the rf potential minimum. The resonant driving force heats the ion, causing a reduction in its fluorescence rate due to its, large, associated Doppler shifts. A typical micro-motion detection scan is presented in fig. \ref{figmicromotion}. The 2D gray scale plots are the ion fluorescence rate as a function of the dc electrode compensation voltage and the drive frequency for both increasing (left) and decreasing (right) frequency sweeps. The asymmetry and hysteretic behavior are a result of the nonlinearity of the trapping potential together with large amplitude of the ion motion\cite{duffing}.

At the optimal compensation voltage of $-0.1$ V, no drop in fluorescence is observed. The linear shift of the resonance as a function of the compensation voltage is expected since compensation is performed only on one electrode and thus slightly modified the secular frequencies. We have observed that the optimal compensation voltages do not vary significantly in time. This points to geometrical imperfections as the source of the ion displacement whereas light or thermally induced charging of dielectric surfaces seems to be negligible \footnote{When the $405$ nm beam directly hits the alumina surface a clear charging effect is observed.}. We found this method for micro-motion compensation to be experimentally simple as compered with the fluorescence modulation method (also know as rf-photon correlation) which we had previously employed. One drawback is the need to scan the drive frequency as the trap frequency is modified by the compensation voltages and, therefore, might make this method more time consuming. However, this problem can be mitigated by first calibrating the dependence of the trap frequency on the compensation voltage and then adjusting the drive frequency accordantly by computer.

\section{Single qubit manipulation}
\label{QubitManipulation}
There are several ways to encode a qubit using the electronic states of a trapped-ion. In the $^{88}$Sr$^+$ ion, which lacks a nuclear spin but has a low lying meta-stable D level, two options can be used. One is encoding the qubit in the $S_{1/2}$ and $D_{5/2}$ levels which are separated by an optical transition. The second ,which is discussed here, is encoding the qubit in the two Zeeman sub-levels of the $S_{1/2}$ ground state.

\subsection{Qubit initialization and detection}
Initialization of the Zeeman qubit is performed by optical pumping, in which a circularly polarized laser beam parallel to the magnetic field ($\sigma^{+/-}$) and in resonance with the $S_{1/2}\rightarrow P_{1/2}$ transition, excites only one of the Zeeman states. A $3~ \mu$s, $\sigma^+$ polarized, pulse initializes the qubit to the $m=+1/2$ spin state with probability greater than $0.99$. The initialization fidelity in this case depends on the polarization purity which at low magnetic field is limited by magnetic field fluctuations and at a high magnetic field by stress-induced birefringence in the vacuum chamber view-port. To improve on the initialization fidelity we further preform optical pumping using the narrow quadrupole transition, $S_{1/2}\rightarrow D_{5/2}$, in which spectral selectivity rather than polarization allows for the excitation of only one of the Zeeman states. Here, errors as small as $10^{-4}$ were obtained \cite{anna_high_fidelity}.

In contrast to initialization, state detection presents a much harder challenge. Since the two qubit levels are typically separated by an energy difference which is smaller than the $P_{1/2}$ excited state linewidth, direct state selective fluorescence is impossible. To discriminant between them, one of the qubit states is shelved to the meta-stable $D_{5/2}$ level prior to detection by resonance fluorescence. The selective shelving of only one spin state is done using the narrow linewidth $674$ nm laser. High fidelity state detection, therefore, imposes stringent requirements on the laser performance. Robust shelving can be achieved by applying several consecutive excitation pulses to different Zeeman levels in the $D_{5/2}$ state manifold. We have demonstrated a total initialization and detection fidelity of $0.9989(1)$ \cite{anna_high_fidelity} which is compatible with recent estimates for the fault-tolerance error threshold.

\subsection{Qubit rotations and coherence}
\begin{figure}[h]
\center
\includegraphics[width=8cm]{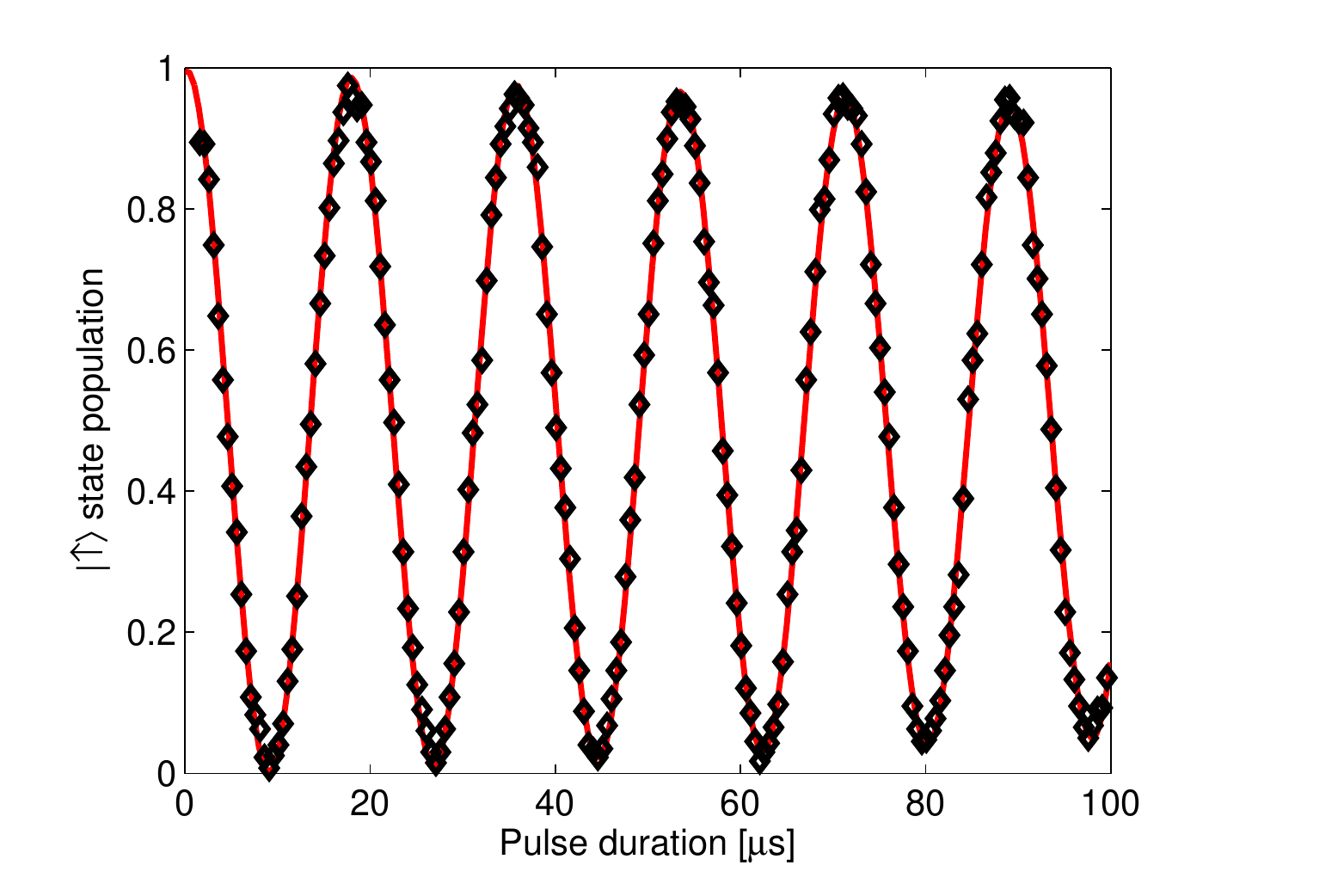}
\caption{Rabi Oscillation between the two state of the Zeeman qubit. The transition is driven by an oscillating magnetic field in resonance with the qubit Zeeman splitting. The red line is a fit to a sin with decaying amplitude}
\label{ZeemanRabi}
\end{figure}

The transition between the two $S_{1/2}$ Zeeman states is driven by resonant oscillating magnetic field. It is produced by a current flowing in one of the rod electrodes that are positioned $1.66$ mm away from the ion; see Fig \ref{trap zoom}(a). The electric current source is an amplified DDS which is impedance matched to a few Ohms by a Balun transformer. The oscillating current has an amplitude of $100$ mA at a frequency range of $1-30$ MHz, and gives rise to a Rabi frequency around $50$ kHz. Rabi oscillations between the two qubit levels are shown Fig.\ref{ZeemanRabi}.

\begin{figure}[h]
\center
\includegraphics[width=8cm]{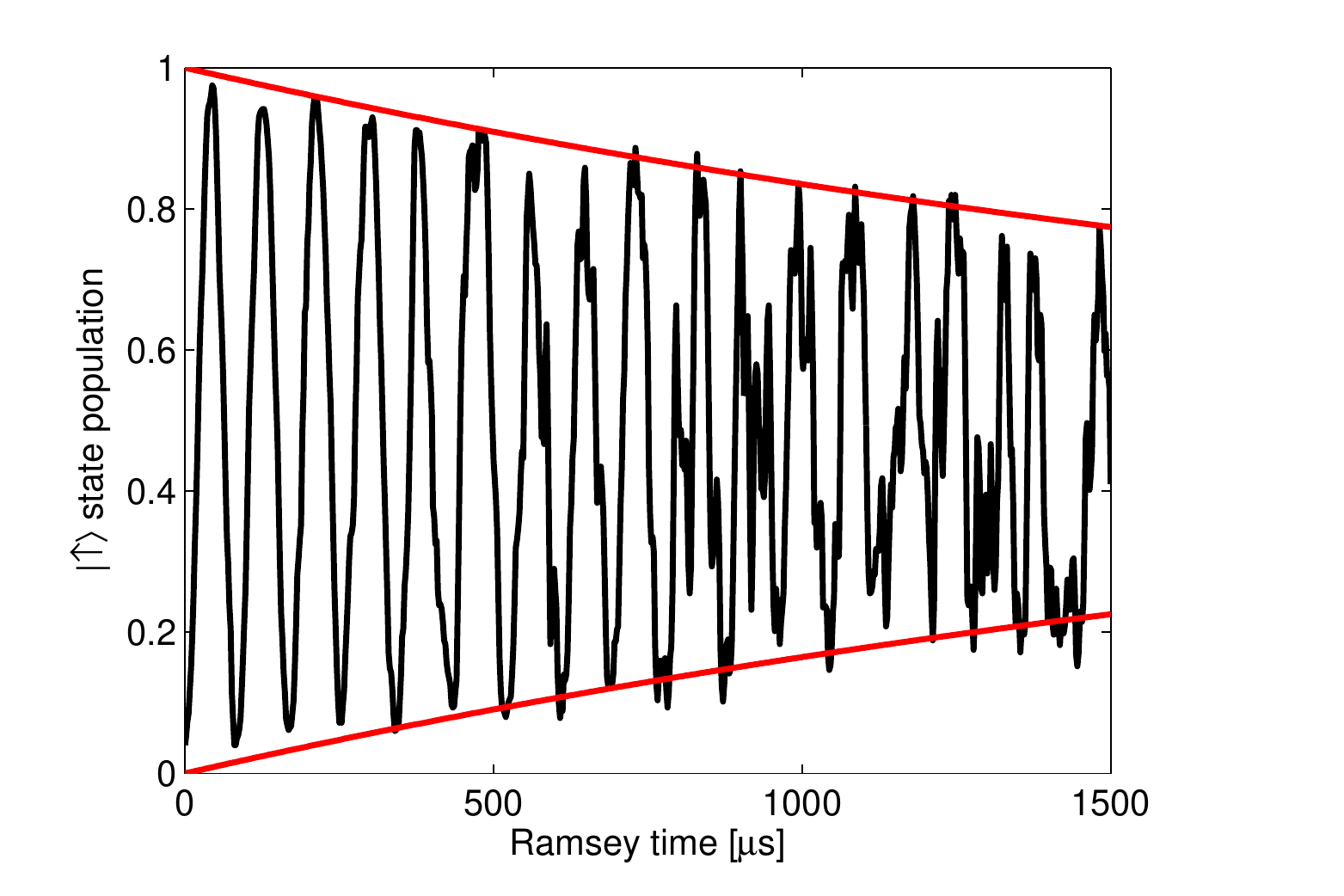}
\caption{Qubit coherence measured is a Ramsey experiment. A small detuning is responsible for the oscillation in the population. A coherence time of 2.5ms is found from an exponential fit to the decaying envelope. The disorder in the oscillation after $500 \mu$s is related to low frequency magnetic noise that does not average in the time needed to take a single point in the figure.}
\label{ZeemanRamsey}
\end{figure}

The coherence of a Zeeman superposition is quickly lost in the presence of magnetic field noise. To protect our qubit from decoherence due to magnetic field noise, we have designed and built a servo system that stabilized the magnetic field on the ion and reduced noise components, mainly at the power line frequency ($50$ Hz) and its harmonics. We were thus able to reduce these magnetic field noise amplitudes to the few $\mu$G level. The coherence time of the qubit, as was measured in a Ramsey experiment, is $2.5$ ms. Coherence oscillations in a typical Ramsey spectroscopy experiment are shown in Fig.\ref{ZeemanRamsey}. Here the coherence time was limited by low frequency ($<1$ Hz) and large amplitude ($>100\mu$G) magnetic field noise, as suggested by the un-smooth appearance of the fringes at Ramsey times longer than $\sim 500\ \mu$s. By further using dynamic decoupling methods our qubit coherence time was extended up to $1.4$ s \cite{Quantum_lockin}.

\section{Cooling and heating rate}
\label{Cooling}

Cooling the ion to a low mean vibrational quantum number, $\bar n$, is useful for both quantum information and precision  measurement applications. Doppler laser cooling is a simple and efficient mechanism for cooling but typically yields $\bar n>1$ . A multi-level cooling cycle, such as in Sr$^{+}$, raises additional complications on reaching the Doppler limit. The atomic line profile on which we preform Doppler cooling is shown in Fig.\ref{IonPic_SPline}(b). We slightly detune the $1092$ nm repump laser red of resonance to create a pronounced dark resonance (seen as a deep in the graph). This results in a steeper slope on which we detune the cooling laser to reach a lower final temperature.

\begin{figure}[h]
\center
\includegraphics[width=8cm]{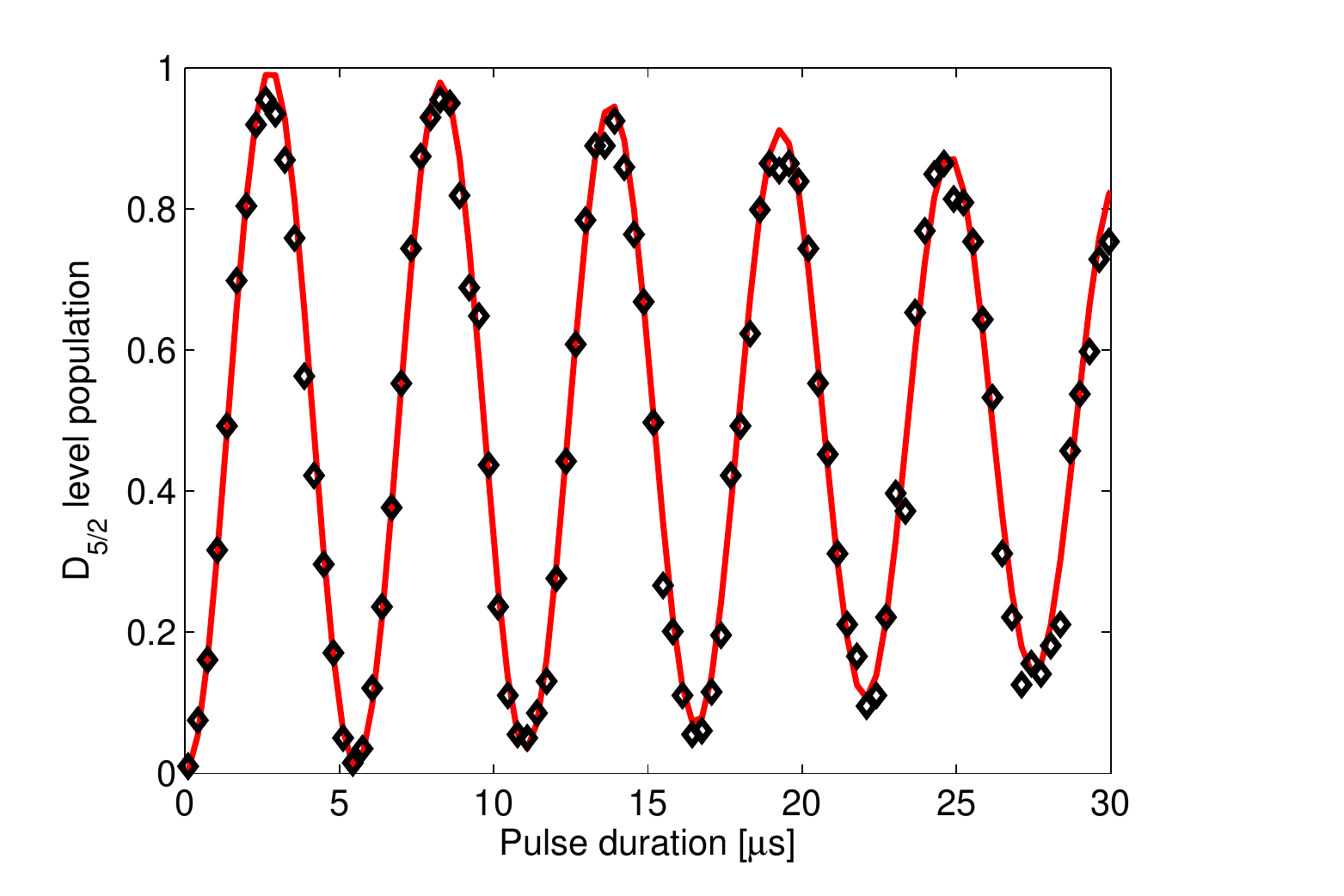}
\caption{Rabi oscillation on the $S_{1/2}\rightarrow D_{5/2}$ carrier transition after Doppler cooling. The red line is a fit to a thermal distribution model which agree with an average vibration occupation number $\bar n = 12$. This value is consistent with the Doppler limit. }
\label{Dooplercooling}
\end{figure}

The ion temperature can be measured in several ways \cite{Wineland_GSC,Epstein2007,Wesenberg2007}. The spectrum of the narrow $S_{1/2}\rightarrow D_{5/2}$ transition has resolved sidebands. The ratio of red to blue sideband excitation probability can be used for thermometry,  when $\bar n \leq 1$. However, after only Doppler cooling $\bar n \gg 1$. Instead, a simple way to estimate the ions' temperature is to measure the decay of the carrier Rabi oscillation due to the ion motion. The carrier Rabi frequency depends on the occupation of the vibrational mode as $\Omega_n\propto1-\eta^2 n$, up to second order in $\eta$. Hence, a thermal distribution leads to different Rabi frequencies which result in oscillation with a decaying envelope. Fig.\ref{Dooplercooling} shows Rabi oscillations on the carrier transition with an average Rabi frequency of $\Omega_0/2\pi=180$ kHz. Fitting the data to a thermal distribution model, shown by the red solid line, yields $\bar n = 12$, consistent with the Doppler cooling limit.

\begin{figure}[h]
\center
\includegraphics[width=8 cm]{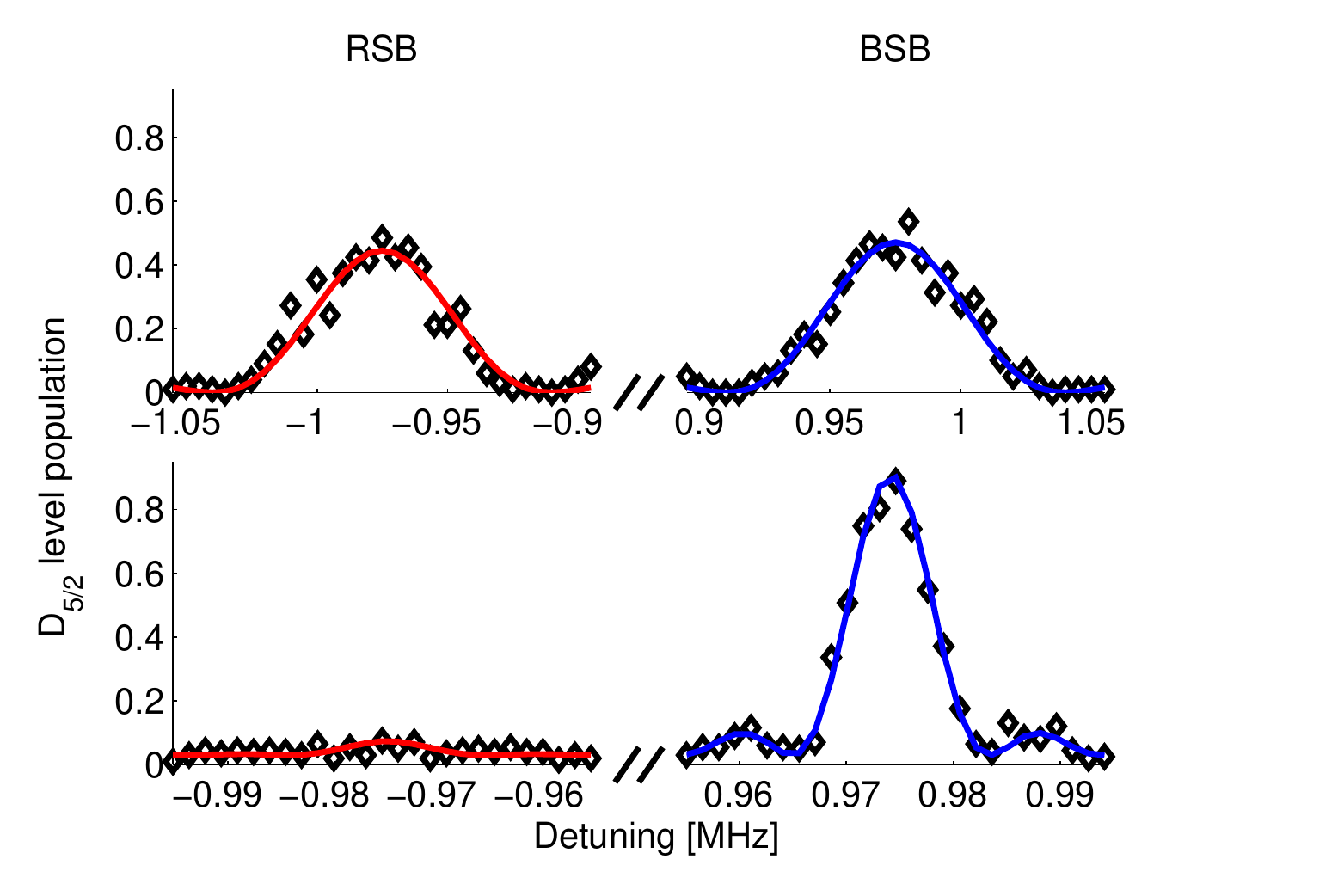}
\caption{Sideband spectroscopy and ground state cooling. (a) just after Doppler cooling and (b) after Doppler and sideband cooling. After sideband cooling, the imbalance between the red and blue sidebands indicates $95\%$ occupation of ground state }
\label{GSC}
\end{figure}

High fidelity qubit operations require $\bar n$ below the Doppler limit and close to the ground state. This is achieved by applying sideband cooling on the narrow $S_{1/2}\rightarrow D_{5/2}$ transition. Since the meta-stable $D_{5/2}$ level is long lived, it needs to be quenched by the $1033$ nm repump laser. By properly choosing the saturation parameter of the quenching laser, continuous sideband cooling can be preformed on the, effectively broadened, $S_{1/2}\rightarrow D_{5/2}$ transition. For optimal performance, the repump laser intensity has to be modified as the ion is cooled, or the cooling time has to be increased \cite{GSC_3level_ion}. In order to shorten the total cooling time, we preform such continuous cooling for only $2$ ms, after which the ion is cooled to $\bar{n} \lesssim 1$. Since most of the population that is not in the ground state is in $n=1$, we then apply two discrete population transfer pulses on the red sideband followed by a $1033$ nm repump pulse. Figure \ref{GSC} shows spectroscopy of the axial red sideband (RSB) and blue sideband (BSB) after (a) Doppler cooling only. Here the sideband excitation pulse duration is $15 \mu$s (b) Resolved sidebands cooling and a pulse duration of $100 \mu$s. Without sideband cooling, the population transfer on the two sidebands is similar, while after sideband cooing, the RSB almost completely disappears. For a thermal distribution model the probabilities to excite the RSB and BSB are given by,
\[P_{ex}^{\{r,b\}}=\sum_n P_{th}(n)\sin^2(\Omega^{\{r,b\}}_nt).\]
Here $P_{th}(n)=\frac{1}{\bar n+1}(\frac{\bar n}{\bar n+1})^n$ is the thermal distribution and $\Omega^{\{r,b\}}_n=\eta_{674}\Omega_0\sqrt{(n+\{0,1\})}$ are the sidebands Rabi frequencies. A fit to the measured data, shown by the blue and red solid lines in Fig.\ref{GSC}, indicates an average occupation number of $\bar n=0.05$. The small background in the measurement is mainly due to carrier excitation resulting from fast frequency noise in the laser spectrum \cite{narrow_laser}. We find this combined protocol of continuous and pulsed cooling to be efficient and robust without the need to dynamically optimize the quenching laser intensity.

\begin{figure}[h]
\center
\includegraphics[width=8cm]{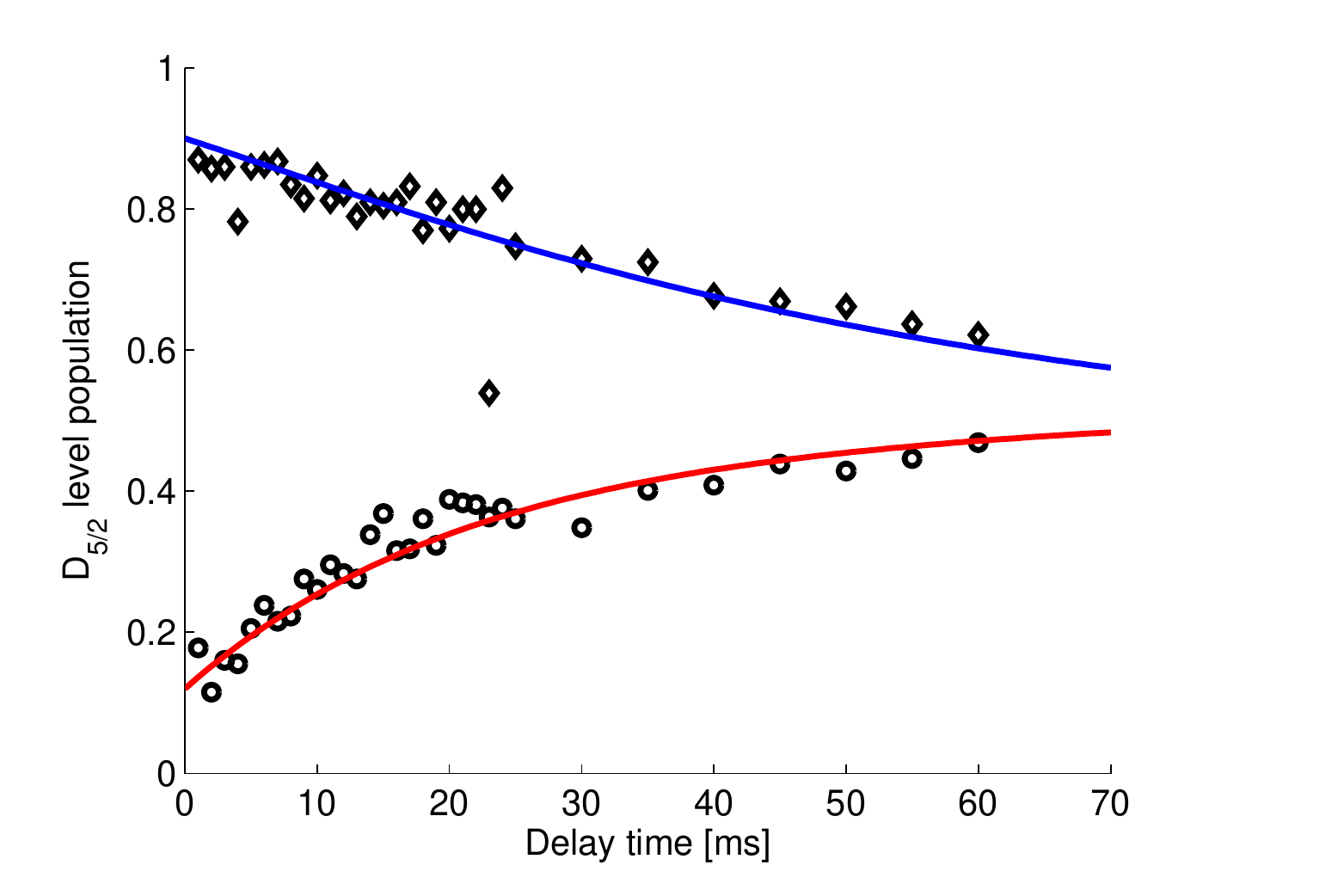}
\caption{Heating rate measurement. Probability to excite the axial RSB (circles) and BSB (diamonds), after sideband cooling, and as a function of time. The solid lines are a fit to a model, assuming a thermal distribution and a constant heating rate. The fit also includes a small constant offset due to the off-resonance excitation of the carrier. The measured heating rate is $\dot{\bar n}=0.016$ ms$^{-1}$.}
\label{heating rate}
\end{figure}

We determined the trap heating rate by varying the time between resolved sideband cooling and the measurement of $\bar n$. Figure \ref{heating rate} shows the excitation probability of the RSB and BSB as function of this delay time. Assuming a thermal distribution and a constant heating rate we fit our data. The fit, shown by the red and blue solid lines, indicates a heating rate of $\dot{\bar n}=0.016$ ms$^{-1}$. Using this measurement we calculate the electric field noise spectral density, $S_E=4m_{Sr}\hbar\omega_{ax}\dot{\bar n}_{ax}/e^2$ \cite{Electric_noise_density}. Assuming, $1/f$ noise, a good quantity to compare between traps of similar dimensions is $\omega S_E$. We measure $\omega S_E=  1.3\cdot10^{-6}$ (V/m)$^2$. This value is consistent with heating rate measurements in traps of similar size \cite{heating_rate_haffner}. During more then two years of operation we did not observe a significant change in the heating rate.

\section{Summary}
In conclusion, we have constructed various basic building blocks necessary for quantum information related experiments with trapped ions. A, sub-mm scale, trap was manually assembled using conventional and laser machining and commercially available parts. The trap harmonic frequencies are in the MHz range, allowing for operation in the Lamb-Dicke regime. All the necessary laser sources were constructed and frequency locked to stable references. A qubit was encoded into the Zeeman splitting of the electronic ground-state of a single $^{88}$Sr$^+$ ion. A narrow linewidth laser enables electron shelving on the $S_{1/2}\rightarrow D_{5/2}$ transition prior to state selective fluorescence detection. Using the red sideband of this transition we have also demonstrated ground-state cooling. Coherent transitions between the qubit states are performed by oscillating a current through an additional electrode which is integrated in the trap structure. By servo stabilization of magnetic field noise, we demonstrate a long qubit coherence time.

\bibliographystyle{nature}
\bibliography{TrapReferences}

\end{document}